\documentclass[fleqn,usenatbib]{mnras}

\usepackage[T1]{fontenc}

\DeclareRobustCommand{\VAN}[3]{#2}
\let\VANthebibliography\thebibliography
\def\thebibliography{\DeclareRobustCommand{\VAN}[3]{##3}\VANthebibliography}

\usepackage{graphicx}	
\usepackage{amsmath}	
\usepackage{newtxtext,newtxmath}

\newcommand{\rbir}{R$_{\rm birth}$}

\newcommand{\rh}{R$_{\rm half}$}
\newcommand{\ith}{$^{\rm th}$}
\newcommand{\dm}{$\Delta$M/M}
\newcommand{\feh}{[Fe/H]}
\newcommand{\fehdisp}{[Fe/H]$_{\rm disp}$}
\newcommand{\sigmarbir}{$\sigma_{\rm Rbirth}$}

\newcommand{\sigmafeh}{$\sigma_{\rm [Fe/H]}$}

\title[Reliability and limitations of inferring birth radii]{Reliability and limitations of inferring birth radii in the Milky Way disk}

\author[Lu et al.]{
Yuxi(Lucy) Lu,$^{1, 2}$,\thanks{E-mail: lucylulu12311@gmail.com}
Tobias Buck$^{3}$, 
Ivan Minchev$^{3}$,
Melissa K. Ness$^{1, 4}$
\\
$^{1}$Department of Astronomy, Columbia University, 550 West 120\ith\ Street, New York, NY, USA\\
$^{2}$American Museum of Natural History, Central Park West, Manhattan, NY, USA\\
$^{3}$Leibniz Institute for Astrophysics Potsdam, An der Sternwarte 16, 14482 Potsdam, Germany\\
$^{4}$Center for Computational Astrophysics, Flatiron Institute, 162 5\ith\ Avenue, Manhattan, NY, USA\\
}

\date{Accepted XXX. Received YYY; in original form ZZZ}

\pubyear{2022}

\begin{document}
\label{firstpage}
\pagerange{\pageref{firstpage}--\pageref{lastpage}}
\maketitle

\begin{abstract}
Recovering the birth radii of observed stars in the Milky Way is one of the ultimate goals of Galactic Archaeology. 
One method to infer the birth radius and the evolution of the ISM metallicity assumes a linear relation between the ISM metallicity with radius at any given look-back time. 
Here we test the reliability of this assumption by using 4 zoom-in cosmological hydrodynamic simulations from the NIHAO-UHD project. 
We find that one can infer precise birth radii only when the stellar disk starts to form, which for our modeled galaxies happens $\sim$ 10 Gyr ago, in agreement with recent estimates for the Milky Way. 
At later times the linear correlation between the ISM metallicity and radius increases, as stellar motions become more ordered and the azimuthal variations of the ISM metallicity start to drop.
The formation of a central bar and perturbations from mergers can increase this uncertainty in the inner and outer disk, respectively.
\end{abstract}




\section{Introduction}
It is widely accepted that stars move away from their birth locations over time in the process commonly referred to as radial migration.
Radial migration was first suggested by \cite{Grenon1972, Grenon1989}, in which they found old metal-rich stars in the solar neighborhood that exhibited kinematic and abundance properties consistent with those of the inner disk.
Stars can move away from their birth radii either because they have high eccentricities or due to a permanent change in their angular momentum.
The latter process is also referred to as radial migration or mixing.
Stars oscillating around their guiding radii can blur the intrinsic relations, however, this process can be accounted for by estimating the stellar guiding radius.
On the other hand, migrating stars are largely indistinguishable in their kinematics from stars born locally \citep[e.g.][]{Sellwood2002, Roskar2008, Minchev2010}.
Moreover, they can flatten the intrinsic abundance gradient in the disk significantly over time \citep[e.g. see Figure 5 from][]{Minchev2013}. 

As a result, to truly understand the enrichment history of the Milky Way (MW), we need to connect the present-day observables to the birth properties of stars.
Without such knowledge, reconstructing MW's formation history remains unreachable, and one can easily infer an incorrect enrichment history of the Galaxy. One example, pointed out in \cite{Minchev2018}, is if we attempt to infer the temporal evolution of the ISM metallicity gradient with stars in mono-age bins, assuming age is equivalent to look-back time, we would end up with a flatter ISM metallicity gradient evolution compared to the truth.

The most ambitious approach for recovering common birth environments of stars is perhaps that of chemical tagging \citep[][]{Freeman2002}. 
This method assumes stars that were born in the same cluster are chemically homogeneous, and that clusters can differentiated from one another given sufficiently precise and numerous individual abundance measurements from different nucloesynthetic channels. 
Then, by grouping stars chemically, one can trace the stars back to their original birth aggregates in the Galaxy. 
However, this method is a challenge for the current data 
\citep[e.g.][]{BlandHawthorn2010, Ness2018, DeSilva2007, Manea2022, Nelson2021,Ting2021,Ratcliffe2020, Ness2022}. 
Nevertheless, even with the currently available abundance measurements, we can still hope to obtain a subset of birth properties of stars with sensible assumptions, one of which is the birth radius of a star.

\cite{Minchev2018} presented a largely model-independent approach for estimating both stellar birth radii, \rbir, and the Galactic ISM metallicity evolution with time and radius, [Fe/H](t, R), using metallicity and age estimates from the
local HARPS sample \citep{Adibekyan2012}. The method relied on the following assumptions: (1) the gas is well mixed azimuthally, (2) stars are born from a narrow metallicity range at a certain radius at any given time, (3) the ISM [Fe/H](t, R) evolved smoothly with both radius and time, and (4) the MW disk formed inside-out. \cite{Feltzing2020} concluded the method shows promising results by inferring \rbir\ with APOGEE DR14 data and comparing the resulted \rbir\ distributions to those in a hydrodynamic simulation from \cite{Agertz2021}.
Other derivations of the radial migration strength can be found in the works by, e.g., \cite{Frankel2019, Frankel2020, Lian2022}.

Although the method described in \cite{Minchev2018} has shown to be successful, the information about birth radius in the MW is lost, and therefore, the inferred \rbir\ from data cannot be truly validated.
Moreover, inferring a star's birth radius with its metallicity and age requires a correlation between the ISM [Fe/H] and R at the time when the star was born (look-back time = age of the star).
In this letter, we will show that precise \rbir\ can only be inferred when a rotationally supported stellar disk is starting to form.
We use MW-like cosmological zoom-in simulations from NIHAO-UHD \citep{Buck2020, Buck2020b}, in which we have access to the enrichment history and \rbir\ of stars.
We will do this by linking the evolution of ISM metallicity, the orderliness of the gas-particle orbits, mergers, and the existence of a stellar bar with the intrinsic uncertainty on inferring \rbir.

In section~\ref{subsection:poss}, we investigate at what time does the correlation between the ISM metallicity and R is strong enough to infer reliable \rbir.
Once the correlation is established, the dispersion around the metallicity -- R linear relation can still lead to intrinsic uncertainty in inferring \rbir.
This scatter can be created from the azimuthal variation of ISM metallicity caused by satellite infall, the spiral arms, the bar, or possibly other perturbations \citep[e.g.][]{Bellardini2021,Spitoni2019,snchez2020}. 
We examine the connection between the ISM metallicity gradient, azimuthal ISM metallicity variation, the orderliness of the gas-particle orbits, and the uncertainty of inferring \rbir, assuming a linear relation between the ISM metallicity and R at all times in section~\ref{subsection:linear}.
Finally, we discuss the implication for inferring \rbir\ in the MW in section~\ref{sec:MW}.

\section{Data \& Methods}
\subsection{Simulated Data}
The simulated galaxies used in this work were first presented by \citep{Buck2018}. 
They are taken from a suite of high-resolution cosmological hydrodynamical simulations of MW-mass galaxies from the NIHAO-UHD project \citep{Buck2020b}. 
A modified version of the smoothed particle hydrodynamics (SPH) solver GASOLINE2 \citep{Wadsley2017} was used to calculate the simulations from cosmological initial conditions. Star formation and feedback (both energetic and chemical elements) are modeled following the prescriptions in \citep{Stinson2006,Stinson2013}.
The simulations used in this work trace oxygen and iron separately \citep[see][for a recent udate to the chemical enrichment]{Buck2021} and adopted a metal diffusion algorithm between particles as described in \citet{Wadsley2008}.
The parameters for simulations used in this letter are taken from \cite{Buck2020b}.
The numbers in the naming represent the total Dark Matter (DM) halo mass in unites of solar mass.
The time separation between every two consecutive output snapshots is $\sim$ 200 Myr, and all of the galaxies have been integrated until redshift 0. 
For more information on the simulation details and galaxy properties, we refer the readers to \cite{Buck2020b}. 

\section{Results}
To infer the birth radius for a star from its [Fe/H] and stellar age alone, a correlation between the ISM metallicity and R at a look-back time equal to the stellar age of the star is required. 
However, as shown in simulations \citep[e.g.][]{Gurvich2022, Bird2013,Bellardini2022}, observations of high-redshift galaxies \citep[e.g.][]{Ciuca2021, Wisnioski2015,ubler2019,Elmegreen2007} and observations of the MW \citep[e.g.][]{Belokurov2022, Conroy2022}, hot, turbulent gas is likely to be common in early galaxy formation.
This means a correlation between the ISM metallicity and R at early times may not exist.
Similarly, as pointed out in \cite{Minchev2018}, if the MW started with a ﬂat gradient, their method would fail for stars $\gtrsim$ 10 Gyr.

\subsection{When can we infer reliable \rbir\ from stellar [Fe/H] and age?}\label{subsection:poss}
\begin{figure}
    \centering
    \includegraphics[width=0.48\textwidth]{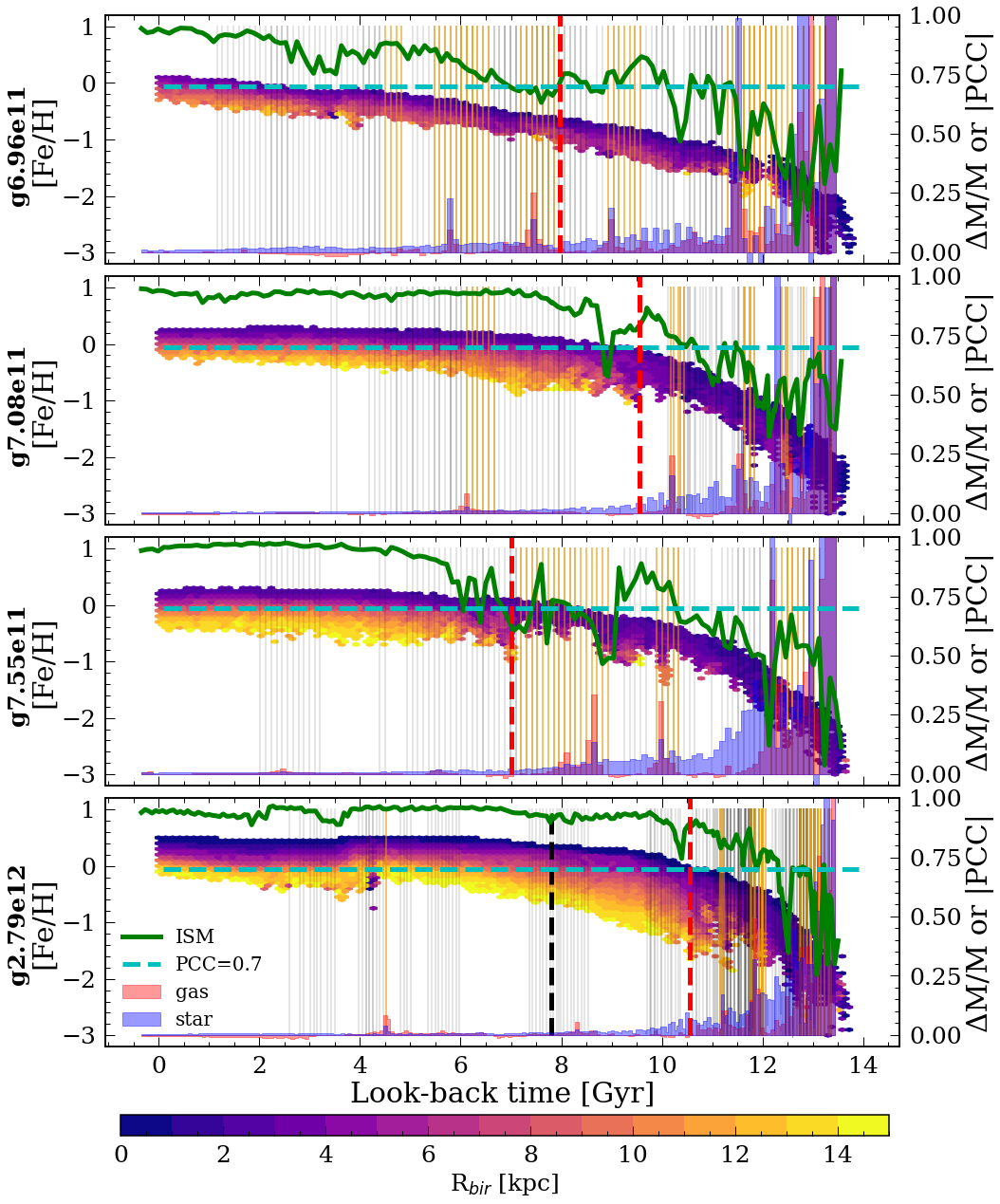}
    \caption{Formation history of 4 galaxies used in this work. 
    The metallicity evolution of the stars is shown in the background, colored by \rbir.
    The blue and red histograms show the instantaneous mass increase (\dm) within 2*\rh\ of the galaxies for star and gas, respectively, where \rh\ is the half-light radius of the galaxy at each look-back time. 
    These values are calculated using (M$_{n+1}$-M$_{n}$)/M$_{n}$, where $n$ is the snapshot number and $n$ = 0 represents the snapshot at a look-back time of 13.8 Gyr. 
    The green lines show the absolute Pearson correlation coefficient (|PCC|) between [Fe/H] and R for the ISM at each look-back time when the snapshot is available.
    The cyan dashed lines indicate PCC = 0.7, which is when the ISM [Fe/H] and R are strongly correlated. 
    The red dashed lines show the approximate times when the stellar disks started to from.
    The black dashed line in the bottom figure shows the formation of the bar taken from Buck et al. (2018) Figure 10.
    The vertical grey/orange lines are the times when there is a satellite > 1\%/10\% total mass ratio in the main galaxy's viral radius, indicating on-going merger events.
    This figure implies that inferring birth radii with \feh and age is most accurate after the turbulent phase of galaxy formation has ended, and when the stellar disk has begun to form, which for the MW is suspected to be $\sim$ 10 Gyr ago based on combining observations and simulations \citep{Belokurov2022}. }
    \label{fig:feh}
\end{figure}

Figure~\ref{fig:feh} summarizes the formation histories of the 4 simulated galaxies of interest.
The metallicity evolution of the stars is shown in the background, colored by \rbir. 
The blue and red histograms show the instantaneous mass increase (\dm) within 2\rh\ of the galaxies for star and gas particles, respectively, where \rh\ is the half-light radius of the galaxies at each look-back time, calculated by (13.8 Gyr - cosmic time). 
\dm\ is calculated by \dm\ = (M$_{n+1}$-M$_{n}$)/M$_{n}$, where $n$ is the snapshot number and $n$ = 0 represents the snapshot at a look-back time of 13.8 Gyr. 
The increase in stellar mass is a combination of in-situ star formation and ex-situ star accretion.
Similarly, the gas mass increase results from in-situ supernova and ex-situ gas accretion.
The red dashed line shows the time when the stellar disk is starting to form, which is defined here as most stars are on circular orbits \citep[see Figure 10 of ][or Figure~\ref{fig:3} left plot]{Buck2020b}.
The black dashed line in the bottom figure shows the formation of the bar taken from \cite{Buck2018} Figure 10.
The vertical grey/orange lines are the times when there is a satellite > 1\%/10\% total mass ratio in the main galaxy's viral radius, indicating on-going merger events.
The red dashed lines, black dashed lines, and the grey/orange lines are used throughout the figures in this letter.
The green line shows the absolute value of the Pearson correlation coefficient (|PCC|) between the metallicity and R of the gas particles at each look-back time when the snapshot is available. 
PCC is a statistical quantity that identifies the linear correlation between two variables, in which -1 and 1 represent a perfectly negative or positive linear correlation, respectively. 
Two strongly correlated variables typically have a |PCC| $>$ 0.7, meaning $\sim$ 50\% of the variation has been accounted for in the data by fitting a line.
We have indicated the line of |PCC| = 0.7 in the Figures (cyan dashed line).

It is clear that all 4 galaxies formed bottom-up, meaning they originated from a series of galaxy mergers and gas accretion events, as the increase in star and gas mass are $>$ 100\% early on (see blue/red histograms).
Most stars are born in the inner galaxy early on, while later on the birth radius distribution broadens, indicating inside-out formation.
During this early time, when \dm\ is large, the |PCC| between the ISM metallicity and R are mostly $<$ 0.7, meaning no strong linear correlation exists, and thus, we cannot infer precise \rbir\ from a star's metallicity and age alone.
The correlation between the ISM metallicity and R remains low until a well-ordered stellar disk begins to form and when a significant amount of stars are formed in the outer disk ($> \sim$ 5 kpc; indicated by the red dashed line) for most galaxies.
However, once the frequent mergers/accretions have subsided and a well ordered stellar disk begins to form, the |PCC| stays above 0.7 for all galaxies\footnote{Due to the frequent merger history at low redshift, galaxy g6.96e11 has a lower |PCC| compared to the other galaxies after its stellar disk has begun to form. }.
This implies that inferring birth radii with \feh and age is most precise and accurate after the turbulent phase of galaxy formation has ended, which for the MW has been suspected to be $\sim$ 10 Gyr ago based on comparing simulations with observations \citep{Belokurov2022}\footnote{\cite{Conroy2022} also pointed out the thick disk formed $\sim$ 13 Gyr ago defined by star formation efficiency. However, for the purpose of this study, we find the definition of disk formation from \cite{Belokurov2022} fits best with this study.}. 

\subsection{ISM metallicity gradient and azimuthal variation, circularisation of the orbits, and the uncertainty on \rbir}\label{subsection:linear}
\begin{figure}
    \centering
    \includegraphics[width=0.45\textwidth]{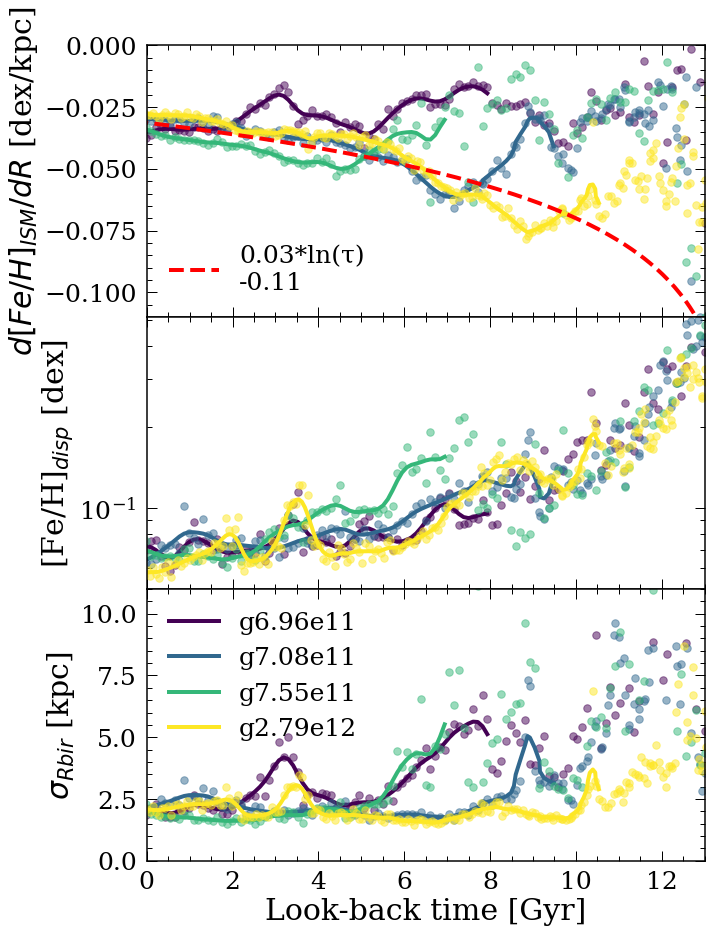}
    \caption{Metallicity gradient evolution (top), metallicity dispersion around the linear fit, \fehdisp\ (middle), and uncertainty in inferring \rbir($\sigma_{Rb}$) assuming the metallicity gradient evolution is known and that there is no error on [Fe/H] and age measurements (bottom).
    The uncertainty, $\sigma_{Rb}$, is calculated by \fehdisp/(|$d$[Fe/H]$_{ISM}$/$d$R|). 
    The solid lines are the smoothed version of the points, using a 1D Gaussian filter, after the stellar disk begins to form.
    The red line in the top plot shows a natural log function to guide the eye.}
    \label{fig:slope}
\end{figure}

How well does a straight line fit the metallicity gradient at all times, and what uncertainty will that bring in inferring \rbir?
The answer lies within Figure~\ref{fig:slope}.
The top plot shows the metallicity gradient evolution over time, and the solid lines display the smoothed version of the points, using a 1D Gaussian filter after the disk has started to form for each galaxy.
The slopes are calculated by fitting a line between the gas particles' metallicity and R within 25 kpc of the galaxy.
The red dashed line shows a natural log function to guide the eye.
\cite{Minchev2018} derived a natural-log functional form for the metallicity gradient evolution, and it fits well with most of the galaxies soon after their disks have begun to form. 
As pointed out earlier, the g6.96e11 galaxy has frequent mergers/infalls at low redshift, causing a more chaotic ISM evolution.
The metallicity dispersion (\fehdisp) around the linear relation between the ISM metallicity and R at each snapshot is shown in the middle plot.
Similar to the conclusion from section~\ref{subsection:poss}, the ISM is more turbulent at earlier times and thus, the |PCC| between ISM [Fe/H] and R is low, and \fehdisp\ is high. 
However, as the stellar disk started to form, \fehdisp\ settled to $<$ 0.1 dex. 

We can also estimate the intrinsic uncertainty in inferring \rbir, $\sigma_{Rb}$, assuming 1) we know the linear fits of the ISM metallicity at a given look-back time, and that 2) there is no error on the observables ([Fe/H] and age; see the bottom plot of Figure~\ref{fig:slope}).
The intrinsic uncertainty is estimated using $\sigma_{Rb}$ = \fehdisp/(|$d$[Fe/H]$_{ISM}$/$d$R|). 
It is clear that even with perfectly measured observables, the intrinsic scattering around the ISM metallicity evolution, assuming a linear relation between the ISM [Fe/H] and R at all times, yields an uncertainty on the \rbir\ measurement to be $\gtrsim$ 2 kpc for all 4 simulated galaxies.
For the MW, based on the estimation of the current metallicity gradient of $\sim$ -0.05 dex/kpc with a dispersion of 0.1 dex derived from Cepheids \citep{Luck2018}, the intrinsic uncertainty would be \sigmarbir $\sim$ 2 kpc, which is comparable with the simulated galaxies.
However, as pointed out by \citep{Luck2018}, the intrinsic dispersion of 0.1 dex is smaller than the measurement uncertainty of 0.14 dex, implying the actual dispersion can be smaller.
We also note that the metallicity gradient in the Galaxy is still not well constrained \citep[e.g.][]{Spina2022, Eilers2022}
    
\begin{figure*}
    \centering
    \includegraphics[width=0.48\textwidth]{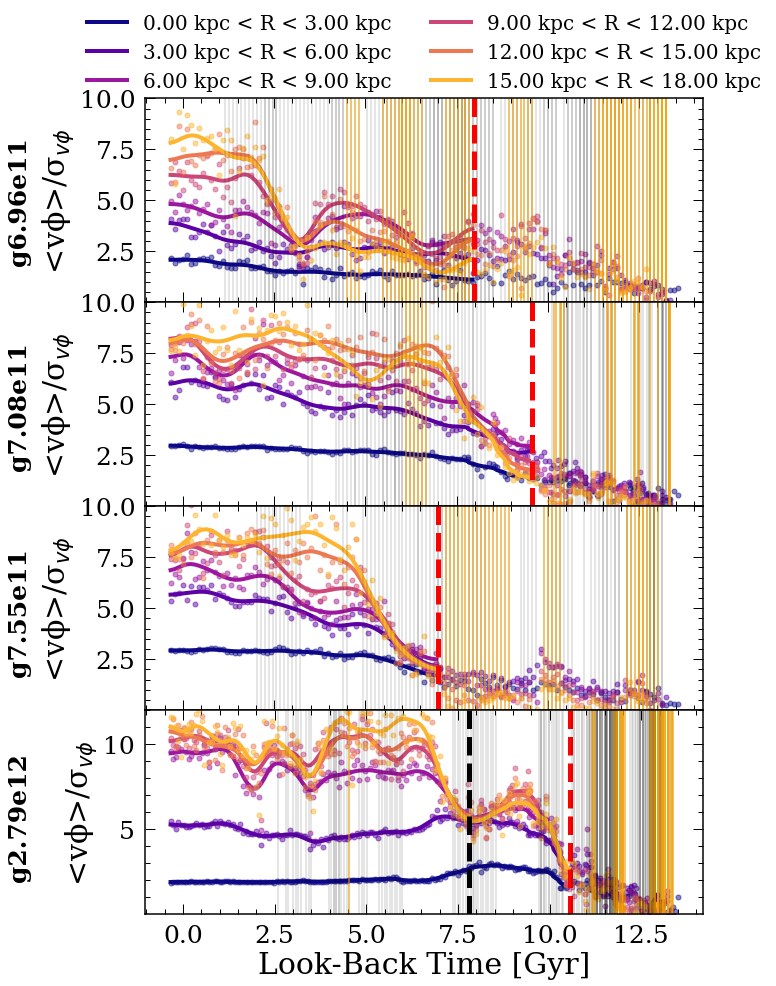}
    \includegraphics[width=0.48\textwidth]{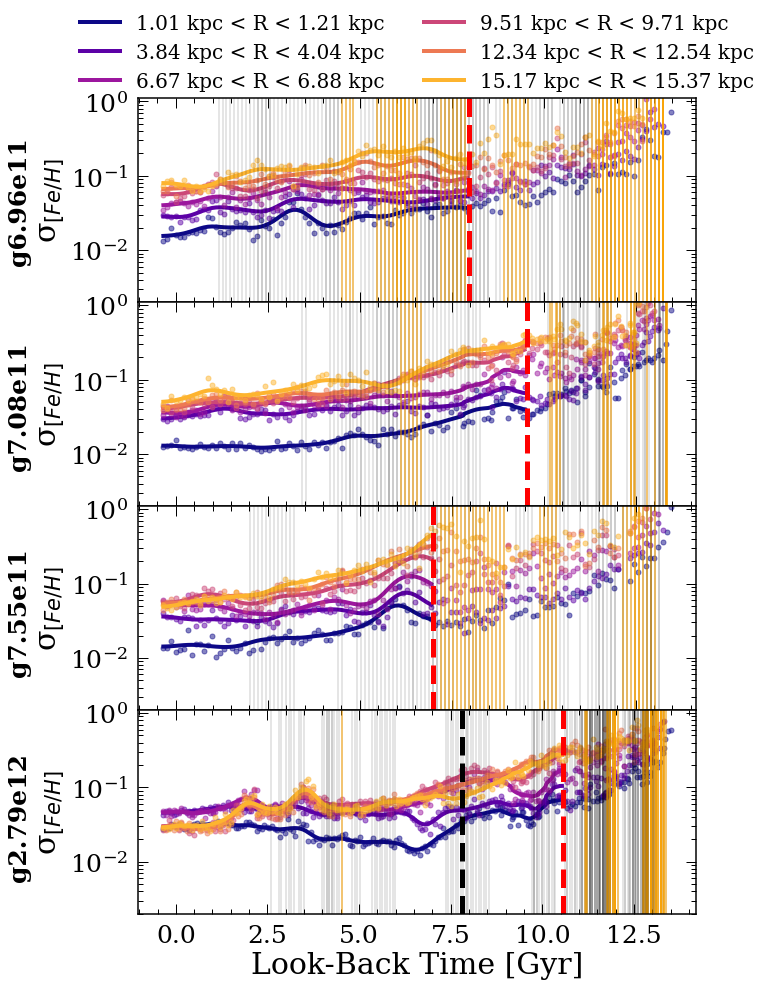}
    \caption{Time evolution of $<v_{\phi}>/\sigma_{v\phi}$ (left) and the azimuthal ISM metallicity variation, \sigmafeh, (right) for the four simulated galaxies, in which $<v_{\phi}>$ is the average azimuthal velocity, and $\sigma_{v\phi}$ is the azimuthal velocity dispersion. 
    $<v_{\phi}>/\sigma_{v\phi}$ can be taken as the ``orderliness'' of the gas-particle orbits, or a proxy for the circularity of the star's orbits.
    We look at $<v_{\phi}>/\sigma_{v\phi}$ in 6 different radial bins and \sigmafeh\ in 6 narrow radial bins ranging from 0-18 kpc.
    The solid lines are the smoothed version of the points, using a 1D Gaussian filter after the stellar disk begins to form. 
    }

    \label{fig:3}
\end{figure*}

We can also connect the ISM evolution and \sigmarbir\ with the ``orderliness'' of the stars orbit (a proxy for circularity of the star's orbits), $<v_{\phi}>/\sigma_{v\phi}$, for the ISM, where $<v_{\phi}>$ is the average azimuthal velocity, and $\sigma_{v\phi}$ is the azimuthal velocity dispersion. 
Figure~\ref{fig:3} left column plots show the time evolution of $<v_{\phi}>/\sigma_{v\phi}$ for the 4 simulated galaxies. 
We look at $<v_{\phi}>/\sigma_{v\phi}$ in 6 different radial bins between 0-18 kpc.
The solid lines are the smoothed version of the points, using a 1D Gaussian filter after the stellar disk begins to form. 
The black dashed line in the bottom figure shows the formation of the bar taken from \cite{Buck2018} Figure 10.
Darker vertical lines mean more satellites exist in the viral radius at that time.
Figure~\ref{fig:3} right plot shows the same as the left but with the ISM metallicty variation in a small radial bin (\sigmafeh) at 6 different radius between 0-18 kpc.
The metallicity variation increases towards the outer disk, and can also be affected by the bar, agreeing with findings from other simulations and chemical evolution models \citep[e.g.][]{Bellardini2021,Spitoni2019}.

Figure~\ref{fig:3} suggest when the stellar disk begins to form, the orbits transition rapidly into a more ordered configuration and the ISM metallicity azimuthal variation also started to decrease.
The rapid increase in $<v_{\phi}>/\sigma_{v\phi}$ and the decrease of ISM metallicity azimuthal variation shows a fast transition from the turbulent gas phase to the quiesent disk phase \citep[see also][]{Gurvich2022}.
This rapid phase transition also leads to a well-established metallicity gradient with a high correlation (|PCC| $>$ 0.7).
This means, after the rapid increase in the orderliness of the gas-particle orbits, a star formed from the ISM has a well-constraint metallicity at a certain radius, meaning we can infer birth radii with smaller \sigmarbir\ (Figure~\ref{fig:slope}). 
However, late mergers can induce turbulence in gas, and therefore, increase \sigmafeh\ and \sigmarbir\ in mostly the outer galaxy, c.f. dispersion in $<v_{\phi}>/\sigma_{v\phi}$.
The bar can also induce variation in metallicity in the inner galaxy. 
As a result, the uncertainty in inferred \rbir\ for the inner Galaxy after the formation of the bar, and the outer Galaxy during a time of merger could be higher. 

\section{Conclusions and Discussion of the implications for the MW}\label{sec:MW}
By using MW-like simulations from the NIHAO-UHD project, we are able to connect the phases of galaxy formation with the possibility of inferring birth radii of stars. In particular, this letter shows:
\begin{itemize}
    \item Precise birth radii can be inferred from stellar ages, guiding center radii, and metallicities only once the stellar disk starts to form - $\sim$ 10 Gyr ago for the MW \citep[][]{Belokurov2022}.
    \item At later times the linear correlation between the ISM metallicity and R increases, as stellar motions become more ordered and the azimuthal variations of the ISM metallicity decrease by an order of magnitude (see Figure~\ref{fig:feh} and Figure~\ref{fig:3}). 
    \item A central bar and  perturbations from mergers can increase the intrinsic uncertainty of inferred \rbir\ in the inner and outer disk by up to 50\%, respectively (see Figure~\ref{fig:3}).
    \item Considering the observed present-day ISM metallicity gradient, derived from Cepheids, of -0.05 dex/kpc with a scatter of 0.1 dex \citep{Luck2018}, the minimum intrinsic uncertainty on inferring \rbir\ in the MW is $\sim$ 2 kpc. 
    However, as pointed out in \cite{Luck2018}, this scatter is smaller than the measurement uncertainty (0.14 dex), and therefor, the true scatter and the intrinsic uncertainty is likely to be smaller than 2 kpc.
\end{itemize}

However, in this study, we only investigated a linear relation between the ISM metallicity and R.
Future work should investigate the possibility that other functional forms exist, that better describe this correlation.
We also only used the ISM gradient derived from Cepheids and other studies have found slightly different gradients ranging from around -0.05 to -0.07 dex/kpc \citep[e.g.][]{Anders2017,Maciel2019}.
Studies derive local metallicity gradient using neutral gas and HII regions have found a slightly shallower gradient of $\sim$ -0.04 dex/kpc \citep[e.g.][]{DeCia2021, Esteban2022}.
Most of the studies either do not report the metallicity intrinsic dispersion around the relations or have a scatter that is on the order of the measurement uncertainty.
As a result, the intrinsic uncertainty on \rbir\ for the MW is not constrained.
We also want to point out that the evolution of the simulated galaxies may not match exactly with that of the MW.
One discrepancy is that the disk seems to form earlier for the MW compared to simulations \citep[see also][]{Belokurov2022}, which could indicate missing/incorrect physics in the simulations.
Future work is needed to further investigate this discrepancy.

\section*{Acknowledgements}
Melissa Ness is supported in part by a Sloan Fellowship. 
Tobias Buck acknowledges support from the European Research Council under ERC-CoG grant CRAGSMAN-646955. 
This research made use of {\sc{pynbody}} \citet{pynbody}.
We gratefully acknowledge the Gauss Centre for Supercomputing e.V. (www.gauss-centre.eu) for funding this project by providing computing time on the GCS Supercomputer SuperMUC at Leibniz Supercomputing Centre (www.lrz.de).
This research was carried out on the High Performance Computing resources at New York University Abu Dhabi. We greatly appreciate the contributions of all these computing allocations.

\section*{Data Availability}
The data underlying this article will be shared on reasonable request to the corresponding author.



\bibliographystyle{mnras}
\bibliography{references} 





\bsp	
\label{lastpage}
\end{document}